\documentclass[aps,nofootinbib,amsmath,prd,twocolumn,showpacs,superscriptaddress,groupedaddress]{revtex4}

\usepackage{graphicx}  
\usepackage{dcolumn}   
\usepackage{bm}        
\usepackage{amssymb}   

\def\calR{{\cal{R}}}

\def\calR{{\cal{R}}}
\def\calG{{\cal{G}}}
\def\hatq{{\tilde{q}}}
\def\hatQ{{\tilde{Q}}}

\def\del{\partial}
\def\vphi{\varphi}
\def\Tr{{\rm Tr}}

\def\MS{{\overline{\rm MS}}}
\def\FRG{{\rm FRG}}
\def\PW{{\rm PW}}
\def\R{{\rm R}}
\def\B{{\rm B}}

\newcommand{\myincludegraphics}[3]{\hspace{-7pt} \mkern10mu \raisebox{- #1 pt}{\scalebox{#2}{\includegraphics{#3}}}}

\begin{document}

\mbox{}

\title{Scheme dependence and universality in the functional renormalization group}

\author{Alessandro Codello}
\email{codello@sissa.it}
\affiliation{SISSA, Via Bonomea 265, 34136 Trieste, Italy}

\author{Maximilian Demmel}
\email{demmel@thep.physik.uni-mainz.de}
\affiliation{PRISMA Cluster of Excellence \& Institute of Physics (THEP),
University of Mainz, D-55128 Mainz, Germany}

\author{Omar Zanusso}
\email{O.Zanusso@science.ru.nl}
\affiliation{Radboud University Nijmegen, Institute for Mathematics, Astrophysics and Particle Physics,
Heyendaalseweg 135, 6525 AJ Nijmegen, The Netherlands}


\begin{abstract}
We prove that the functional renormalization group flow equation admits a perturbative solution
and show explicitly the scheme transformation that relates it to the standard schemes of perturbation theory.
We then define a universal scheme within the functional renormalization group.
\end{abstract}

\maketitle

\section{Introduction.}
\label{intro}

In the functional renormalization group (FRG) approach to quantum field theory (QFT),
the effective average action (EAA) $\Gamma_k[\varphi]$ is a scale-dependent functional
that interpolates between the effective action (EA) $\Gamma[\varphi]$ of a quantum field theory when the scale $k$ is zero $\Gamma_{k=0}[\varphi]=\Gamma[\varphi]$,
and a bare UV action $S_\Lambda[\varphi]$ when $k$ is equal to a given UV scale $\Lambda$.
The scale dependence of the EAA is governed by the flow equation
\begin{eqnarray}\label{erge}
\partial_t \Gamma_k[\vphi]
=
\frac{\hbar}{2}
\Tr \left(\Gamma^{(2)}_k[\varphi]+R_k\right)^{-1}\partial_t R_k\,,
\end{eqnarray}
where we used $t=\log k$ to parametrize the change with the scale $k$ \cite{Wetterich:1992yh}.
In \eqref{erge} we introduced the cutoff $R_k$ that modifies the propagator of the IR modes and that
makes \eqref{erge} both IR and UV finite.
Importantly, the property $R_{k=0}=0$ ensures that the method reproduces the EA of the system.
The integration of the flow to $k=0$ will provide us a renormalized EA and we will indicate renormalized quantities with the subscript ${\rm R}$.

\section{Perturbative solution of the functional renormalization group}

It is possible to provide a solution of \eqref{erge} as a perturbative expansion in powers of $\hbar$ \cite{Litim:2002xm}. We first expand
\begin{eqnarray}\label{bare_expansion}
\Gamma_k[\vphi] = S_{\rm B}[\vphi] + \sum_{L\geq1} \hbar^L \Gamma_{L,\,k}[\vphi]\,.
\end{eqnarray}
The functional $S_{\rm B}$ will play the role of bare action of the method, as will become clear below.
Plugging \eqref{bare_expansion} in \eqref{erge} a flow equation for each order $\Gamma_{L,\,k}[\vphi]$ can be derived by comparing powers of $\hbar$ on both sides.
The first three orders are
\begin{equation}
\begin{split}
 \label{bare_flows}
 \partial_t S_\B[\vphi] &= 0\,,\\
 \partial_t \Gamma_{1,\,k}[\vphi] &= \frac{1}{2} \Tr \left( G_{\B,\,k} \partial_t R_k \right) \,,\\
 \partial_t \Gamma_{2,\,k}[\vphi] &=
 \frac{1}{2} \Tr \left( \Gamma^{(2)}_{1,\,k}[\vphi]\, \partial_t G_{\B,\,k} \right) \,,
\end{split}
\end{equation}
where we defined a modified propagator
$$G_{\B,\,k}\equiv \left(S^{(2)}_\B[\varphi]+R_k\right)^{-1}\,.$$
Each flow equation of this system can be separately integrated in $k$ by showing that the right-hand side is a total $t$ derivative.
The procedure, however, requires regularization because commuting the operators $\partial_t$ and $\Tr$ spoils the UV finiteness of the result.
We thus regularize the functional trace
\begin{equation}\label{commutator}
\Tr \,\partial_t = \partial_t \,\Tr_{\rm reg}\,.
\end{equation}
Any known regularization technique can be applied to \eqref{commutator}.
In the following, we will adopt dimensional regularization to make the closest contact with the standard methods of perturbation theory.
Integrating the first and second orders we obtain
\begin{equation}
\begin{split}
\label{bare_functionals}
\Gamma_{1,\, k}[\vphi]
&=
\frac{1}{2} \Tr_{\rm reg}\,{\log \left(S^{(2)}_{\rm B}[\vphi]+R_k\right)}\,,
\\
\Gamma_{2,\, k}[\vphi]
&=
-\frac{1}{12}\,\myincludegraphics{10}{0.15}{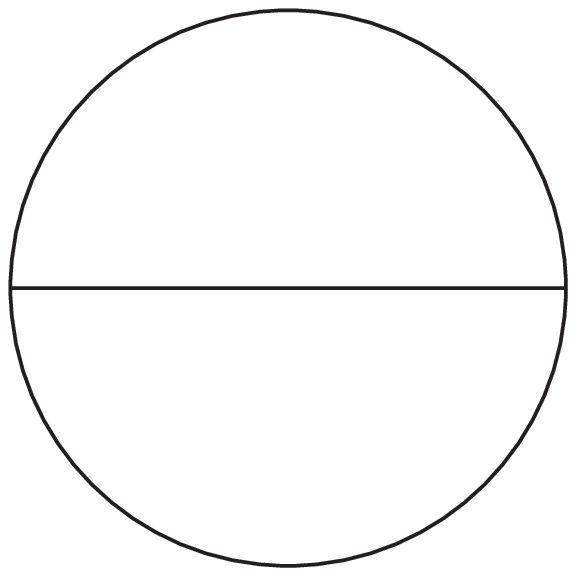}
+\frac{1}{8}\,\myincludegraphics{10}{0.15}{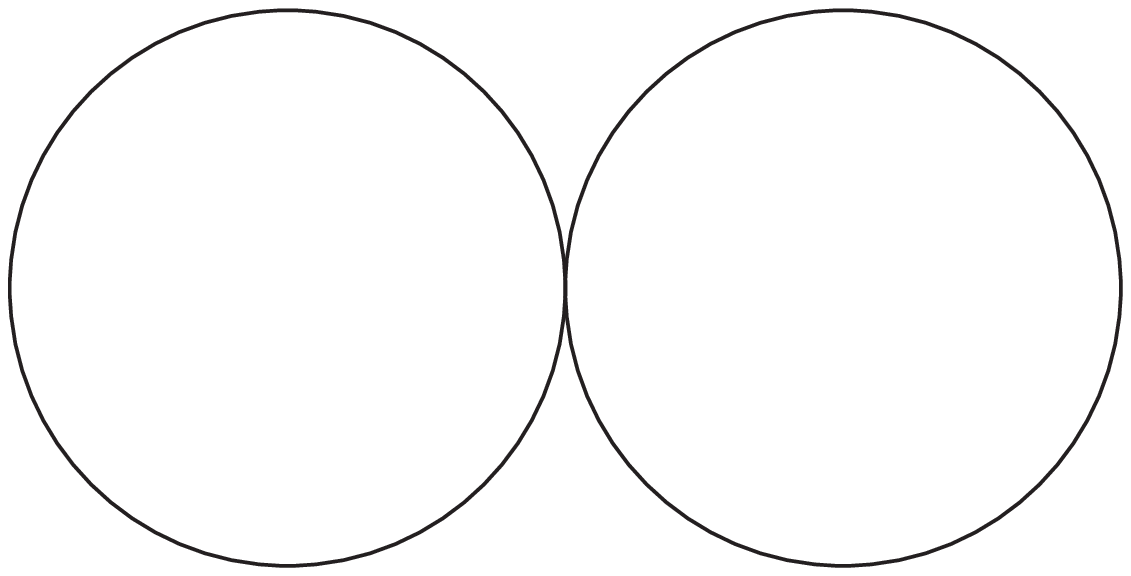}\,,
\end{split}
\end{equation}
where the lines represent $G_{\B,\,k}$.

The traces \eqref{bare_functionals} are regularized by analytically continuing $d$ to the value $d=d_c-\epsilon$,
where $d_c$ is the upper critical dimension and $\epsilon>0$ but small: divergences appear as poles of the form $1/\epsilon^L$ \cite{'tHooft:1972fi}.
The divergences have to be canceled by a suitable renormalization technique.
We thus introduce a further expansion consisting of a renormalized tree-level action $S_{\rm R}$ and counterterms
to subtract the divergences
\begin{eqnarray}\label{counterterms}
S_{\rm B}[\vphi] = S_{\rm R}[\vphi] + \sum_{L \geq 1} \hbar^L \delta S_{L}[\vphi]\,.
\end{eqnarray}
The counterterms $\delta S_L[\vphi]$ have to cancel the divergences of $\Gamma_{L,\,k}[\vphi]$ order by order in $\hbar$
using the prescription of the $\MS$ method that introduces a reference scale $\mu$ \cite{'tHooft:1972fi}.
We choose
\begin{eqnarray}\label{counterterms_choice}
\delta S_{L} \equiv - \Gamma_{L,\, k}^{\rm div} = - \Gamma_{L,\, k=0}^{\rm div}\,.
\end{eqnarray}
The crucial assumption of \eqref{counterterms_choice} is that the divergences are not dressed by the scale $k$.
Equation~\eqref{counterterms_choice} ensures that bare and renormalized vertices share in form the same expansion.
It is possible to prove in general that \eqref{counterterms_choice} holds \emph{within} the formalism,
giving thus a consistency check of the method. We will discuss the implications of \eqref{counterterms_choice} in the example below,
but it could be interesting to speculate on violations of \eqref{counterterms_choice}. We shall leave the possibility open.

\section{Scalar $\phi^4$ model in $d=4-\epsilon$ dimensions}
\label{sect2}

To illustrate the technique of solving the FRG perturbatively,
it is convenient to resort to the example of a scalar $\phi^4$ model in proximity of the critical dimension $d=4-\epsilon$.
The bare action of the theory is
\begin{eqnarray}\label{bare_action_scalar}
 S_{\rm B} [\vphi ]
 =
 \int\!{\rm d}^d x\;
 \!\left\{ \frac{1}{2}\left( \del_\mu\vphi\right)^2
  + \frac{m_\B^2}{2} \vphi^2 +\frac{\lambda_\B }{4!} \vphi^4 \right\}.
\end{eqnarray}
We parametrize the renormalized action as in perturbation theory
\begin{eqnarray}\label{renormalized_action_scalar}
 S_{\rm R} [\vphi_\R ]
 =
 \int\!{\rm d}^d x\;
 \!\left\{ \frac{1}{2}\left( \del_\mu\vphi_\R\right)^2
  + \frac{m_\R^2}{2} \vphi_\R^2 +\frac{\lambda_\R\mu^\epsilon }{4!} \vphi_\R^4 \right\},
\end{eqnarray}
where we introduced the renormalized field $\vphi_\R\equiv Z_\B^{-1/2}\vphi$ and renormalized couplings $m_\R$ and $\lambda_\R$.
The counterterms $\delta S = \sum_{L \geq 1} \hbar^L \delta S_{L}$ are parametrized as
\begin{equation}
\begin{split}\nonumber
 \int\!{\rm d}^d x\;
 &
 \!\Bigl\{ \frac{\delta Z_\B}{2}\left( \del_\mu\vphi_\R\right)^2
  + \frac{m_\R^2 \delta Z_m}{2} \vphi_\R^2
 +\frac{\lambda_\R\mu^\epsilon \delta Z_\lambda }{4!} \vphi_\R^4 \Bigr\},
\end{split}
\end{equation}
and, through \eqref{counterterms}, define implicitly the renormalization constants $Z_\B=1 + \delta Z_\B $, $Z_m=1+\delta Z_m$ and $Z_\lambda=1+\delta Z_\lambda$
as functions of bare and renormalized couplings.
These constants renormalize the theory canceling the divergences of the Feynman diagrams through \eqref{counterterms_choice}.

As easily evinced from \eqref{bare_functionals}, up to two loops
the diagrams involved in the renormalization of \eqref{bare_action_scalar} are the same involved in the standard perturbation theory.
The only difference lies in the propagator that is here modified by the IR cutoff $R_k$ and in momentum space takes the form
$(q^2+m^2_R+R_k(q^2))^{-1}$.
For the computation, we found it convenient to choose the optimized form \cite{Litim:2000ci} given by
\begin{equation}\label{optimized}
 R_k(q^2) = (k^2-q^2)\theta(k^2-q^2)\,.
\end{equation}
We illustrate the effects of the IR cutoff taking a closer look at two diagrams.
At one loop, the relevant diagram for the computation of $Z_\lambda$ is
\begin{eqnarray}\label{1loop_diagram}
 -\frac{3}{2}\, \myincludegraphics{10}{0.15}{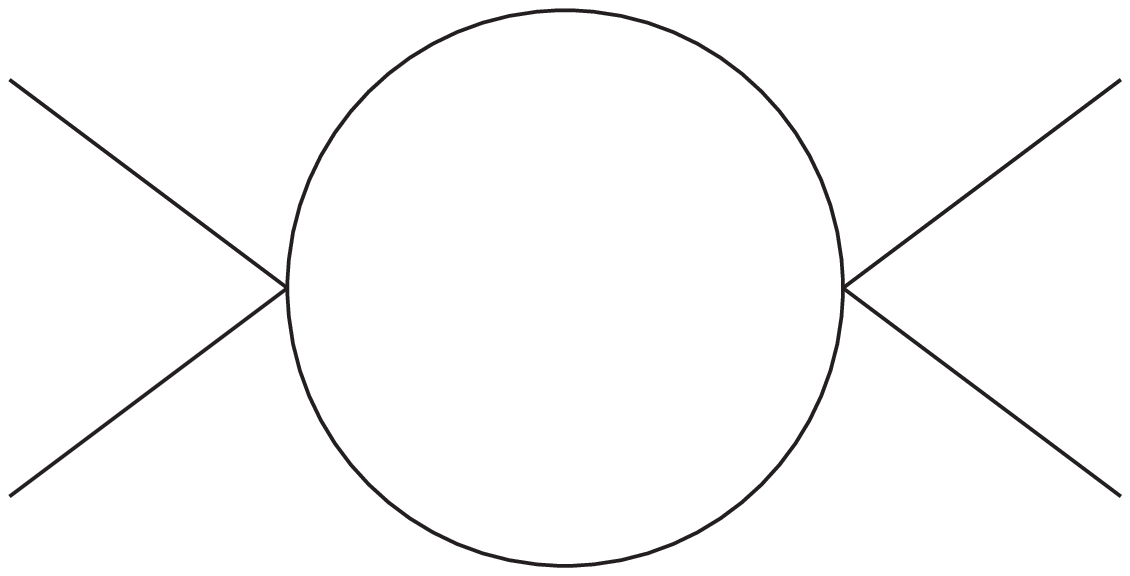}
 &=
 -\frac{3\lambda_\R^2}{16\pi^2\epsilon}
 +\frac{3\lambda_\R^2}{32 \pi^2}\log \!\left(\frac{k^2+m_\R^2}{\mu^2}\right)
 \nonumber\\
 &+
 \;3\frac{2 m_\R^4+2 k^2 m_\R^2-k^4}{64 \pi^2\left(k^2+m_\R^2\right)^2}\!\lambda_\R^2+{\cal O}\!\left(\epsilon\right)\!,
\end{eqnarray}
whose finite part we gave at zero external momenta.
At two loops one of the relevant diagrams is
\begin{eqnarray}\label{2loop_diagram}
\myincludegraphics{8}{0.09}{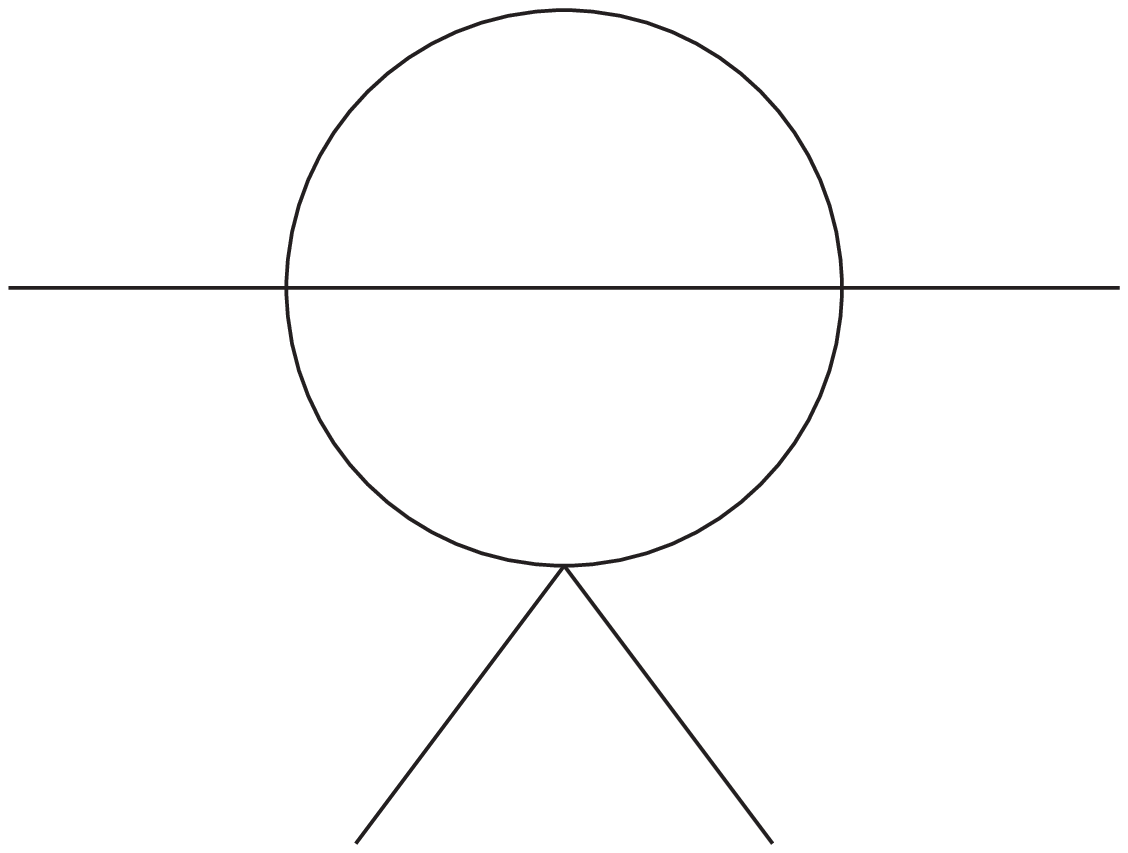}
&=&
\frac{\lambda_\R^3}{128\pi^4\epsilon^2}
-\frac{\lambda_\R^3}{256 \pi ^4 \epsilon }
-\frac{\lambda_\R^3}{128 \pi ^4 \epsilon }\log \!\left(\frac{k^2+m_\R^2}{\mu^2}\right)
 \nonumber\\
&+&
 \frac{\left(2 m_\R^2 +3 k^2\right)k^2}{256 \pi ^4 \epsilon \left(k^2+m_\R^2\right)^2}\lambda_\R^3
 + {\rm finite}.
\end{eqnarray}
The diagrams \eqref{1loop_diagram} and \eqref{2loop_diagram} illustrate the general property
that the highest divergence of the $L$th loop, which diverges as $1/\epsilon^{L}$, is \emph{never} dressed by the FRG scale $k$.
This property is not hard to be shown to hold for any admissible cutoff choice.
Subdivergences and finite parts, instead, do depend on $k$.
However, counterdiagrams appearing from second loop order on are tailored to cancel against subdivergences.
Therefore, it is possible to prove in general that the divergent part of the EAA does not depend on $k$
consistently with the requirement \eqref{counterterms_choice}.
The renormalization constants take the standard values of the $\MS$ scheme
\begin{eqnarray}
 \delta Z_m = \frac{\lambda_\R }{16 \pi^2 \epsilon }
 + \frac{\lambda_\R^2}{128 \pi^4 \epsilon^2}
 -\frac{\lambda_\R^2}{512 \pi^4 \epsilon };
 &&
 \delta Z_\B = -\frac{\lambda_\R^2}{3072 \pi^4 \epsilon }
\nonumber\\
 \delta Z_\lambda = \frac{3 \lambda_\R }{16 \pi^2 \epsilon }
 + \frac{9\lambda_\R^2}{256 \pi^4 \epsilon^2}
 -\frac{3\lambda_\R^2}{256 \pi^4 \epsilon }.
 &&
\end{eqnarray}
The renormalization is completed requiring that $\lambda_\B$ is independent of the reference scale $\mu$ as a function of $\lambda_\R$ \cite{ZinnJustin:2002ru}.
We can derive the $\beta$ function and the anomalous dimension in the usual way
\begin{equation}\label{universal_betas}
\begin{split}
 \beta
 &= \frac{\partial\lambda_\R(\mu)}{\partial\log\mu}
 =
 -\epsilon \lambda_\R + \frac{3\lambda_\R^2}{16 \pi^2}-\frac{17\lambda_\R^3}{768 \pi^4}
 \\
 \eta
 &= \beta(\lambda_\R)\frac{\partial\log Z_\B}{\partial \lambda_\R}
 =
 \frac{\lambda_\R^2}{1536\pi^4}\,.
\end{split}
\end{equation}
The above results, together with the diagrammatic expansion arising from \eqref{bare_functionals},
show that the FRG method reproduces perturbation theory and, in particular, the $\MS$ scheme at two loops.
It is however instructive to elaborate further in this direction.
The scale $k$ of the FRG plays a crucial role in reconstructing the loop expansion \eqref{bare_functionals}
and unveils the presence of divergences that required renormalization through \eqref{commutator}.
However, the property \eqref{counterterms_choice} allows one to subtract the divergences, and therefore to renormalize,
using solely the reference scale $\mu$ of the $\MS$ scheme, while $k$ plays no role.
This implies that the renormalized coupling $\lambda_\R$ defined in \eqref{renormalized_action_scalar}
is actually the renormalized coupling of the $\MS$ scheme and therefore we shall refer to it as $\lambda_\MS=\lambda_\R$,
while its $\beta$ function will be $\beta_\MS=\beta(\lambda_\MS)$.
Generally, the couplings of the FRG method are defined through an operatorial expansion of the EAA of the form
\begin{equation}\label{truncation}
 \Gamma_k[\varphi] = \int\! {\rm d}^dx \sum_i g_i(k) {\cal O}_i\!\left(\varphi\right)\,,
\end{equation}
where $g_i$ are the $k$-dependent couplings, ${\cal O}_i\!\left(\varphi\right)$ are the corresponding operators
and the index $i$ ranges over all possible operators compatible with the symmetries of the system ($\varphi$ parity in the case of the simple scalar \cite{Codello:2012sc}).
The FRG beta functions of the couplings $g_i$ are defined as their $t$ derivatives and computed by inserting \eqref{truncation} in \eqref{erge}.
We define $\lambda_\FRG$ as the coefficient of the $\varphi^4/4!$ operator
or, alternatively, the local part of the four-point function of the model.
Using the perturbative technique described above we have access to an expression for $\lambda_\FRG$
as a function of $\lambda_\MS$ through the finite parts of the loop expansion.
The relevant result at one loop is contained in the finite part of \eqref{1loop_diagram} as
\begin{eqnarray}\label{pre_scheme}
\lambda_\FRG(k)
&=&
\lambda_\MS(\mu)+\frac{3}{32 \pi ^2} \!\log \!\left(\frac{k^2+m_\R^2}{\mu^2}\right)\lambda_\MS^2(\mu)\nonumber
\\
&+& \frac{3}{64 \pi^2} \frac{2 m_\R^4+2 k^2 m_\R^2-k^4}{\left(k^2+m_\R^2\right)^2}\lambda_\MS^2(\mu) .
\end{eqnarray}
This computation can be performed for any coupling $g_i$ and at any loop order
(in particular beyond the one loop method developed in \cite{Ellwanger:1997tp}).
Specializing \eqref{pre_scheme} to the case $k=\mu$ we derive a perturbative scheme-change relation
between the couplings of the $\MS$ and FRG schemes
\begin{equation}\label{scheme}
\begin{split}
\lambda_\FRG(\mu)
&=
\lambda_\MS(\mu)+{\cal F}\!\left(\frac{m^2_\R}{\mu^2}\right)\lambda_\MS^2(\mu)
 +{\cal O}\!\left(\lambda^3\right)\!.
\end{split}
\end{equation}

The transformation \eqref{scheme} is a function of the dimensionless ratio of the renormalized mass with the RG scale because of dimensional reasons,
and its form depends on the $\FRG$ scheme through the choice we made for the IR cutoff \eqref{optimized}.
A similar result was obtained in \cite{Liao:1992fm}, where an equivalent relation between the $\MS$ scheme and the Wilsonian blocking RG is computed.
Presumably, the result \eqref{scheme} may reduce to what is obtained in \cite{Liao:1992fm}
if a specific sharp cutoff $R_k$ is chosen in place of \eqref{optimized} to mimic the effects of blocking.
The method outlined in this section, however, is very general
and the computation can be repeated for any cutoff choice and for any perturbative regularization method at will and at any loop order.

It is instructive to compute the beta function of the $\FRG$ scheme \emph{from} the one of the $\MS$ scheme.
At one loop we obtain
\begin{equation}\label{beta_FRG}
\begin{split}
 \beta_\FRG
 &= \beta_\MS - 2  \frac{m^2_R}{\mu^2}{\cal F}'
 \lambda_\MS^2
 +2{\cal F}
 \lambda_\MS \beta_\MS
\\
 &= \frac{3 \lambda_{\rm FRG}^2 \mu^6}{16 \pi^2 \left(\mu^2+m^2\right)^3}
\end{split}
\end{equation}
where at this order the mass can belong to either scheme and the perturbative inverse of \eqref{scheme} was used.
The result \eqref{beta_FRG} is in agreement with the computations performed in the FRG approach \cite{Codello:2012sc}.
Furthermore $\beta_\FRG$ underlies the fact that the beta functions of the FRG approach are not universal in the customary sense of QFT.
This is due to the fact that the FRG method is a mass-dependent scheme and manifests through the nontrivial coupling of mass and scale in \eqref{scheme}.
Nevertheless it is possible to explicitly and perturbatively map the results of $\MS$ and FRG.

\section{The Papenbrock-Wetterich scheme}
\label{sect3}

The topic of universality of FRG results is discussed in \cite{Pernici:1998tp,Bonanno:1997dj,Arnone:2003pa} and is treated extensively in \cite{Rosten:2010vm}.
We shall now outline the construction of a new scheme, first hinted by Papenbrock and Wetterich in \cite{Papenbrock:1994kf},
that obtains universal results within the FRG method.
A truncation of \eqref{truncation} containing all the operators that are generated at one loop
and that contributes to the flow of the local part of the four-point function of the scalar model is
\begin{eqnarray}\label{truncation_PW}
 \Gamma_k[\varphi]
 &=&
 \int \!{\rm d}^4x \Bigl\{\frac{Z}{2} \left(\del_\mu\varphi\right)^2
 + g_2 \varphi^2
 + g_4 \varphi^4
 + \varphi^2 f_1(\Delta) \varphi^2
 \nonumber\\
 &&
 + g_6 \varphi^6
 + f_2(\Delta_1,\Delta_2,\Delta_3) \varphi_1^2\varphi_2^2\varphi_3^2\Bigr\}.
\end{eqnarray}
We introduced three couplings $g_{2,4,6}$ and two form factors $f_{1,2}$ which contain an amount of information equivalent to infinitely many couplings.
The notation for the second form factor $f_2$ is understood as follows: each Laplacian $\Delta_j=-\partial^2_{x_j}$
acts only on the corresponding insertion $\varphi^2_j=\varphi^2(x_j)$
and subsequently the limit $x_1=x_2=x_3=x$ has to be taken.
The form factors $f_{1,2}$ resemble closely those of the nonlocal heat-kernel expansion \cite{Codello:2012kq}
and satisfy the boundary conditions $f_1(0)=0$ and $f_2(0,0,0)=0$ to have unambiguous definitions of $g_4$ and $g_6$.
All couplings and form factors implicitly encode the scale dependence $k$ which is driven by the flow \eqref{erge}.
Introducing the renormalized field $\varphi_R = Z^{1/2}\varphi$,
we define anomalous dimension $\eta$, dimensionless renormalized couplings $\tilde{g}_{2,4,6}$, and form factors $\tilde{f}_{1,2}$ in momentum space
\begin{equation}
\begin{split}
 g_2 = Z k^2 \tilde{g}_2\,, \quad & \eta = -\partial\log Z/\partial \log k\,,\\
 g_4 = Z^2 \tilde{g}_4\,, \quad & f_1 (q^2) = Z^2 \tilde{f}_1 ( q^2/k^2)\,,\\
 g_6 = Z^3 k^{-2} \tilde{g}_6\,, \quad & f_2 (q_i^2) = Z^3 k^{-2} \tilde{f}_2 (q_i^2/k^2)\,.
\end{split}
\end{equation}
Inserting \eqref{truncation_PW} in \eqref{erge} we compute the FRG beta function $\partial_t\tilde{g}_4$ and $\eta$,
respectively, from the coefficient of the local part of the four-point function
and the order $p^2$ of the two-point function with incoming momentum $p^\mu$ as outlined in \cite{Vacca:2010mj}.
The leading result as a function of all other couplings is
\begin{eqnarray}\label{beta_and_eta}
 \partial_t \tilde{g}_4 &=&
 2 \eta \tilde{g}_4
 + 72 \tilde{g}_4^2 \int_\hatq \calG_\hatq^3 {\rm d} \calR_\hatq
 -432 \tilde{g}_4^2 \tilde{g}_{2} \int_\hatq \calG_\hatq^4 {\rm d} \calR_\hatq
 \nonumber\\
 &&
 + 96 \tilde{g}_4 \int_\hatq \calG_\hatq^3 {\rm d} \calR_\hatq\; \tilde{f}_{1}(\hatq^2)
 - 15 \tilde{g}_{6} \int_\hatq \calG_\hatq^2 {\rm d} \calR_\hatq
 \nonumber\\
 &&
 - 8 \tilde{g}_4^3
 \int_\hatq \calG_\hatq^2 {\rm d} \calR_\hatq \; \tilde{f}_{2}(\hatq^2,-\hatq^2,0)\,,
 \\
 \eta &=&
 -8 \int_\hatq \calG_\hatq^2 {\rm d}\calR_q \tilde{f}_{1}'(\hatq^2)
 -4 \int_\hatq \calG_\hatq^2 {\rm d}\calR_q \hatq^2\tilde{f}_{1}''(\hatq^2)\,,
 \nonumber
\end{eqnarray}
where we introduced a dimensionless momentum integration $\int_\hatq = (2\pi)^{-d}\int\! {\rm d}^4 \hatq$ in $\hatq=q/k$,
that is the natural argument of dimensionless renormalized propagator $ \calG_{\tilde{q}} = Z k^2( Z q^2+R_k)^{-1}$
and derivative of the cutoff $ {\rm d}\calR_{\tilde{q}} = Z^{-1}k^{-2} \partial_t R_k$ in momentum space.
The results \eqref{beta_and_eta} show that $\partial_t \tilde{g}_4$ and $\eta$ only depend on the other dimensionless renormalized couplings as expected on dimensional grounds.
The difference between \eqref{beta_and_eta} and the universal results \eqref{universal_betas}
is that the former underlies a flow that requires the inclusion of potentially infinitely many couplings for consistency,
while the latter depends solely on the coupling $\lambda$.
This difference is the fundamental distinction of the FRG and $\MS$ methods.
While previously we found a dictionary to translate the couplings from one scheme to the other in the form of \eqref{scheme},
we now seek for a consistent closure of \eqref{beta_and_eta} within \eqref{truncation_PW}.
One prescription is obtained first by computing the flow of all the couplings and form factors \emph{but} $\tilde{g}_4$,
and then by setting them at the Gaussian fixed point (GFP) as a function of $\tilde{g}_4$.
We call the one-dimensional submanifold of the theory space obtained in this way generalized GFP (gGFP).
The curve is parametrized by $\tilde{g}_4$ and ends in the GFP when $\tilde{g}_4=0$.
We outline the method with two examples. The leading beta function of $\tilde{g}_2$ and its gGFP are
\begin{equation}\label{gGFP1}\nonumber
\begin{split}
 \partial_t \tilde{g}_2
 = -2 \tilde{g}_2
 - 6\tilde{g}_4 \int_\hatq \calG^2_\hatq {\rm d}\calR_\hatq\,,
 &\quad
 \tilde{g}_{2\,*}
 =
 - 3\tilde{g}_4 \int_\hatq \calG^2_\hatq {\rm d}\calR_\hatq\,,
\end{split}
\end{equation}
and a similar structure holds for $\tilde{g}_6$.
Form factors have flows that can be computed too and the gGFP conditions correspond to differential equations as illustrated from the $\tilde{f}_1(\hatq^2)$ example
\begin{equation}\nonumber
\begin{split}
\partial_t \tilde{f}_1
-2 \eta \tilde{f}_1
-2 \tilde{f}'_1 \hatq^2
&=
72 \tilde{g}^2_4 \int_\hatQ \left( \calG_{\hatQ+\hatq} - \calG_{\hatQ} \right)
 \calG_\hatQ^2 {\rm d} \calR_\hatQ\,,
\end{split}
\end{equation}
that can be solved with the method of characteristics
\begin{equation}\label{gGFP2}\nonumber
\begin{split}
\tilde{f}_{1\,*}(\hatq^2)&= -36 \tilde{g}_4^2 \int_\hatQ (\calG_{\hatQ+\hatq}-\calG_{\hatQ}) \calG_\hatQ\,.
\end{split}
\end{equation}
A similar procedure can be carried over for the other form factor.
It turns out that the gGFP values of the couplings are equivalent to those obtained by directly using the one loop EAA \eqref{bare_functionals},
thus implying that the bare theory underlying \eqref{truncation_PW} is actually massless.
We now define the beta function and anomalous dimension of the Papenbrock-Wetterich scheme (PW)
as the single-coupling beta function that is obtained inserting all the gGFP values in \eqref{beta_and_eta}
\begin{equation}
\begin{split}
 \beta_\PW (\tilde{g}_4)
 &=
 \partial_t
 \tilde{g}_4(\eta,\tilde{g}_{2\,*},\tilde{g}_{4},\tilde{g}_{6\,*},\tilde{f}_{1\,*},\tilde{f}_{2\,*})\,,
\end{split}
\end{equation}
and similarly for $\eta$. We adopt an exponential cutoff for the computation
\begin{equation}\label{exp_cutoff}
 R_k(z) = Z \frac{z}{{\rm e}^{z/k^2}-1}\,,
\end{equation}
and move to the conventional normalization of the coupling $\tilde{g}_4=\lambda_\PW / 4!$.
A new name was adopted for the coupling to underline that it belongs to a new scheme.
The nested integrals appearing in \eqref{beta_and_eta} can be solved analytically along the lines described in \cite{Papenbrock:1994kf,Morris:1999ba}.
The flow $\beta_\PW=\partial_t\lambda_\PW$ is two loops universal
\begin{equation}\label{universal_PW}
\begin{split}
 \beta_\PW
 =
 \frac{3\lambda_\PW^2}{16 \pi^2}-\frac{17\lambda_\PW^3}{768 \pi^4}\,,
 &\quad
 \eta
 =
 \frac{\lambda_\PW^2}{1536\pi^4}\,,
\end{split}
\end{equation}
as seen by comparing with \eqref{universal_betas}.
The coupling $\lambda_\PW$, however, is not $\lambda_\MS$.
In fact, with the techniques developed above we can compute at one loop
\begin{equation}\label{scheme_PW}
\begin{split}
\lambda_\PW(\mu) = \lambda_\MS(\mu) +
\frac{\log 8 - 3\gamma}{32\pi^2} \lambda^2_\MS(\mu)\,,
\end{split}
\end{equation}
with $\gamma$ the Euler-Mascheroni constant.
Equation~\eqref{scheme_PW} differs from \eqref{scheme} because of the cutoff \eqref{exp_cutoff}
and the absence of a bare mass\footnote{The gGFP mass \eqref{gGFP1} is a correction of order $\lambda_\PW$
that would affect \eqref{scheme_PW} starting from the order $\lambda^3_\MS$.
The beta functions \eqref{universal_betas} and \eqref{universal_PW} will then differ from the order $\lambda^4$ on,
in agreement with the fact that three loop results are not universal.},
and preserves the universality of \eqref{universal_betas}.

\section{Conclusions}

Motivated by the desire of bridging a gap that exists between two powerful approaches to quantum field theory,
we proved that the functional renormalization group flow equation admits a perturbative solution
and showed explicitly that this solution can be related to the standard schemes of perturbation theory.
As a reference technique for the perturbative computations we intentionally used the $\MS$ method, being the most well-known and applied technique in phenomenology.

In Sec.~\ref{sect2} we addressed the question of the scheme dependence in the $\FRG$ approach,
which was previously almost never addressed in the literature with very few exceptions,
and provided a scheme transformation between the $\FRG$ and $\MS$ methods.
The transformation is shown to belong to the class of transformations that do not preserve universality of the beta functions,
as we illustrated through the example of a simple scalar field in four dimensions,
and holds in the overlapping region of validity of the $\FRG$ and $\MS$ methods which correspond to the vicinity of a Gaussian fixed point.
The results of Sec.~\ref{sect2} are very similar in spirit to those of \cite{Liao:1992fm},
where the Kadanoff and Wilson's blocking is investigated in the loop expansion
and the relation between blocking and $\MS$ methods is obtained at one loop.
In fact, the discussion made in \cite{Liao:1992fm} on the limitations of the $\MS$ method
and concerning the role of irrelevant operators can be motivated as well by the results of this paper.
The results on the scheme change are expected to prove valuable
when comparing observables of the phenomenologically more interesting Yang-Mills theories \cite{Morris:2005tv}.

In Sec.~\ref{sect3} we rigorously defined the PW scheme
that restores universality of the beta functions in a FRG setting.
It is a nontrivial feature since the FRG method is a mass-dependent scheme and therefore expected to violate two loop universality.
In the PW scheme two loop universality is achieved by considering a truncation of the space of couplings
that includes all operators that are perturbatively generated at one loop,
and thus the method is reminiscent of the results of \cite{Sonoda:1990gp}.
Our results thus help bridge a gap that exists between the methods that use truncations of the effective action,
and those that renormalize perturbatively through the relevant deformations.
The EAA appearing in Sec.~\ref{sect3} is thus a prototype for a truncation
that is capable of providing two loops universal results when dealing with a renormalizable quantum field theory.

\emph{Acknowledgments.}
The authors thank J.~Pawlowski and M.~Reuter for many useful discussions.
The research of O.Z.~is supported by the DFG within the Emmy-Noether program (Grant No. SA/1975 1-1).



\vfill


\begin{thebibliography}{99}

\bibitem{Wetterich:1992yh}
  C.~Wetterich,
  Phys.\ Lett.\ B {\bf 301}, 90 (1993).

\bibitem{Litim:2002xm}
  D.~F.~Litim and J.~M.~Pawlowski,
  Phys.\ Rev.\ D {\bf 66}, 025030 (2002)
  [hep-th/0202188].


\bibitem{'tHooft:1972fi}
  G.~'t Hooft and M.~J.~G.~Veltman,
  Nucl.\ Phys.\ B {\bf 44}, 189 (1972).

\bibitem{Litim:2000ci}
  D.~F.~Litim,
  Phys.\ Lett.\ B {\bf 486}, 92 (2000)
  [hep-th/0005245].

\bibitem{ZinnJustin:2002ru}
  J.~Zinn-Justin,
  Int.\ Ser.\ Monogr.\ Phys.\  {\bf 113}, 1 (2002).

\bibitem{Codello:2012sc}
  A.~Codello,
  J.\ Phys.\ A {\bf 45}, 465006 (2012)
  [arXiv:1204.3877 [hep-th]].

\bibitem{Ellwanger:1997tp}
  U.~Ellwanger,
  Z.\ Phys.\ C {\bf 76}, 721 (1997)
  [hep-ph/9702309].

\bibitem{Liao:1992fm}
  S.~-B.~Liao and J.~Polonyi,
  Annals Phys.\  {\bf 222}, 122 (1993).

\bibitem{Pernici:1998tp} 
  M.~Pernici and M.~Raciti,
  Nucl.\ Phys.\ B {\bf 531}, 560 (1998)
  [hep-th/9803212].

\bibitem{Bonanno:1997dj}
  A.~Bonanno and D.~Zappala,
  Phys.\ Rev.\ D {\bf 57}, 7383 (1998)
  [hep-th/9712038].

\bibitem{Arnone:2003pa} 
  S.~Arnone, A.~Gatti, T.~R.~Morris and O.~J.~Rosten,
  Phys.\ Rev.\ D {\bf 69}, 065009 (2004)
  [hep-th/0309242].
  
\bibitem{Rosten:2010vm}
  \emph{See} Sect.~VI \emph{of}
  O.~J.~Rosten,
  Phys.\ Rept.\  {\bf 511}, 177 (2012)
  [arXiv:1003.1366 [hep-th]].

\bibitem{Papenbrock:1994kf}
  T.~Papenbrock and C.~Wetterich,
  Z.\ Phys.\ C {\bf 65}, 519 (1995)
  [hep-th/9403164].

\bibitem{Morris:1999ba} 
  T.~R.~Morris and J.~F.~Tighe,
  JHEP {\bf 9908}, 007 (1999)
  [hep-th/9906166].

\bibitem{Codello:2012kq}
  A.~Codello and O.~Zanusso,
  J.\ Math.\ Phys.\  {\bf 54}, 013513 (2013)
  [arXiv:1203.2034 [math-ph]].

\bibitem{Vacca:2010mj} 
  G.~P.~Vacca and O.~Zanusso,
  Phys.\ Rev.\ Lett.\  {\bf 105}, 231601 (2010)
  [arXiv:1009.1735 [hep-th]].

\bibitem{Sonoda:1990gp} 
  H.~Sonoda,
  Nucl.\ Phys.\ B {\bf 352}, 585 (1991).

\bibitem{Morris:2005tv} 
  T.~R.~Morris and O.~J.~Rosten,
  Phys.\ Rev.\ D {\bf 73}, 065003 (2006)
  [hep-th/0508026].

\end{thebibliography}
\end{document}